# Improved emission of SiV diamond color centers embedded into concave plasmonic core-shell nanoresonators

András Szenes[1], Balázs Bánhelyi[2], Lóránt Zs. Szabó[1],
Gábor Szabó[1], Tibor Csendes[2], Mária Csete[1,*]

[1]Department of Optics and Quantum Electronics, University of Szeged, Dóm tér 9, Szeged, H-6720, Hungary.
[2]Institute of Informatics, University of Szeged, Árpád tér 2, Szeged, H-6720, Hungary
*mcsete@physx.u-szeged.hu, +36-62-544654

## Abstract

Configuration of three different concave silver core-shell nanoresonators was numerically optimized to enhance the excitation and emission of embedded silicon vacancy (SiV) diamond color centers simultaneously. According to the tradeoff between the radiative rate enhancement and quantum efficiency (*QE*) conditional optimization was performed to ensure ~2-3-4 and 5-fold apparent *cQE* enhancement of SiV color centers with ~10% intrinsic *QE*. The enhancement spectra, as well as the near-field and charge distribution were inspected to uncover the physics underlying behind the optical responses. The conditionally optimized coupled systems were qualified by the product of the radiative rate enhancements at the excitation and emission, which is nominated as $P_x$ factor. The optimized spherical core-shell nanoresonator containing a centralized emitter is capable of enhancing considerably the emission via bonding dipolar resonance. The $P_x$ factor is 529-fold with 49.7% *cQE* at the emission. Decentralization of the emitter leads to appearance of higher order multipolar modes, which is not advantageous caused by their nonradiative nature. Transversal and longitudinal dipolar resonances of the optimized ellipsoidal core-shell resonator were tuned to the excitation and emission, respectively. The simultaneous enhancements result in $6.2 \cdot 10^5$ $P_x$ factor with 50.6% *cQE* at the emission. Rod-shaped concave core-shell nanoresonators exploit similarly transversal and longitudinal dipolar resonances, moreover they enhance the fluorescence more significantly due to their antenna-like geometry. $P_x$ factor of $8.34 \cdot 10^5$ enhancement is achievable while the *cQE* is 50.3% at the emission. The enhancement can result in $2.03 \cdot 10^6$-fold $P_x$ factor, when the criterion regarding the minimum *QE* is set to 20%.

## 1. Introduction

Metal nanoparticles are capable of coupling light into collective electron oscillations at their resonance frequencies, which phenomenon is nominated as localized surface plasmon resonance (LSPR). The LSPR is of great interest due to the accompanying strong EM-field localization into regions significantly smaller than the illuminating wavelength, which can be exploited in different application areas including high sensitivity biosensing and photo-thermal cancer therapy [1, 2]. Metal nanoshells on dielectric cores are interesting plasmonic structures, since they have widely tunable spectral and near-field properties, which have been already described in the literature [3]. Metal nanoshells on their boundary can support primitive sphere- and cavity plasmons, hybridized modes as well as higher order multipoles by leaving the quasistatic limit i.e. when the particle size is smaller than λ/10, where λ is the wavelength in the surrounding medium [4, 5]. It is an intriguing property of core-shell particles that the same dipolar mode can appear in case of different geometries, however these dipolar modes are accompanied by different number of multipoles, and surprisingly larger scattering cross-section can be achieved via thicker shells [4].

Geometrical tuning of different multipolar modes into specific wavelength intervals was realized, and the differences in the far-field patterns were analyzed [5]. The geometrical tuning is ensured by the hybridization of primitive plasmons supported by the core-shell type nano-objects. The simplest hybridization of primitive sphere and cavity plasmons results in a lower energy resonance with a symmetric charge distribution and in a higher energy resonance with an antisymmetric charge separation, which are also referred as bonding and antibonding modes, analogously to the molecular orbital theory [5, 6, 7]. Many works have been presented about, how the energy of these plasmon modes depends on the dielectric properties of the embedded core and the surrounding medium, as well as on the ratio of the core radius and the full composite nanoparticle radius, which is nominated as the generalized aspect ratio (*AR*) [8, 9]. In a homogeneous dielectric medium the energy of a nanoshell plasmon resonance is determined solely by the *AR* in the quasistatic limit [8].

Namely, increase of the *AR* strengthens the interaction between the primitive plasmons, which red-shifts the symmetric mode-, while blue-shifts the antisymmetric mode-related resonance peak. In general, increase of the permittivity of the surrounding medium promotes the appearance of higher order multipolar modes at larger wavelengths and red-shifts the already existing plasmon resonances. Since the plasmon peak shift is linearly proportional to the refractive index modification of the embedding medium, bio-sensing applications have been developed [5, 10]. In core-shell design considerations it is important to note, that an embedding medium with higher permittivity makes the resonances more sensitive to *AR* changes.

Modification of the core's permittivity has similar spectral effect to that of the surroundings', however the symmetric bonding mode is more sensitive to the embedding medium, while the antisymmetric antibonding mode exhibits larger sensitivity to the changes of the core's medium [9]. Interestingly, in case of nanoshells the larger particle radius does not lead essentially to a stronger scattering. It is especially true for the core-shell particles consisting of multiple dielectric and metal layers, e.g. for the so-called nanomatryoskas, where modification of the middle dielectric layer thickness makes it possible to control the scattering efficiency [11, 12]. All these phenomena can be explained based on the hybridization model with some restrictions, and experiments also prove that by modifying the surroundings' and core's dielectric properties or the *AR* of the nanoshell, the plasmon resonances can be tuned from IR through the UV [5-7].

The near-field enhancement accompanying the LSPR on various nanoparticles has important applications. Several works have been devoted to exploit the LSPR to enhance fluorescence of different emitters [13-15]. All these efforts are based on that the absorption and scattering cross-sections of plasmonic nanoresonators can be detuned via a properly designed geometry. Simple metal nanorods are especially promising candidates for fluorescence enhancement due to their co-existent longitudinal and transversal LSPR occurring at wavelengths tunable by the geometry [16]. In our previous study we have shown those precisely tunable properties, which make the noble metal nanorods capable of enhancing the emission and excitation simultaneously [17]. Nanoshells can also support various multipolar resonances due to the plasmon hybridization, which leads to distinct peaks on their extinction spectra [4, 5]. Elongated nanoshells also show distinct plasmon resonances according to the symmetry breaking stem from different axes and to the hybridization of the corresponding primitive modes [18-20]. Moreover, nanoshells with reduced symmetry could also sustain various plasmonic resonances [21-23]. However, still a few studies have been realized to use them as fluorescence enhancing particles [24]. This is caused by the retardation effects and the electron scattering phenomenon, which limits the resonator quality properties of larger core-shell particles [25, 26]. Accordingly, there is a great demand for core-shell resonators capable of approaching the materials related limits.

Diamond color centers are stable single-photon sources and possess unpaired electron spin functioning as a qubit at room temperature hence they are important for quantum information processing (QIP) applications. Among them, silicon vacancy centers (SiV) are the most promising novel diamond defects for QIP and quantum cryptography applications due to their advantageous relaxation time, stability, uniquely narrow spectral lines and the rare orthogonality of dipoles corresponding to the transition moments of excitation and emission [27-29].

In this paper, we present results about the fluorescence enhancement of silicon vacancy diamond color centers via various concave core-shell type nanoresonators. To enhance the fluorescence of SiV centers, diamond cores were embedded into silver nanoshells, and their optical response was optimized via geometry tuning. The purpose was to enhance the radiative rate at the excitation and emission wavelengths simultaneously. The intuitive expectation is that the *QE* of the coupled system could be enhanced with respect to that of the silver nanorod enhanced SiV centers inspected previously, according to the reduced amount of the absorbing metal [17]. Optimization of different types of concave core-shell nano-resonators illumination configuration has been performed to demonstrate their special advantages and to uncover the underlying nanophotonics.

## 2. Methods

Calculations were realized via the commercially available COMSOL Multiphysics RF module, by applying our previously developed optical response readout methodology [17]. The fluorescent color center is approximated by a pure point-like electric dipole embedded into a diamond dielectric environment and the $P^{rad}$ dipole power radiated towards the far-field and the $P^{non-rad}$ power dissipated as a heat inside the metal, acting as an inhomogeneous environment, were calculated. The quantities characterizing the coupled SiV center - core-shell nanoresonator system are the *Purcell factor* of the dipole [30, 31] and *QE* quantum efficiency of the core-shell nanoresonator, as well as the product of them, which equals to the $\delta R^{rad}$ radiative rate enhancement with respect to vacuum, i.e. is identical with the fluorescence rate enhancement [17]:

$$\delta R^{rad} = Purcell \cdot QE = \frac{P^{rad} + P^{non-rad}}{P_0^{rad}} \cdot \frac{P^{rad}}{P^{rad} + P^{non-rad}} = \frac{P^{rad}}{P_0^{rad}}, \quad (1)$$

where $P_0^{rad}$ is the power emitted by the point-like electric dipole in vacuum. During the numerical computations the emitter was treated as a lossless dipole, hence the $QE_0 \sim 10\%$ intrinsic quantum efficiency of SiV centers at their emission wavelength has been taken into account during the post-analysis. Accordingly, the ideal $QE$ has been corrected to determine the corrected $cQE$ at the emission as follows:

$$cQE = \frac{P^{rad}}{P^{rad} + P^{non-rad} + \frac{1-QE_0}{QE_0}}. \quad (2)$$

The $\delta QE$ apparent quantum efficiency enhancement of SiV centers is also presented for each coupled systems, which is the ratio of the $cQE$ corrected quantum efficiency and the intrinsic $QE_0$. According to reciprocity theorem the $\delta R_{excitation}^{rad}$ excitation and $\delta R_{emission}^{rad}$ emission rate enhancement can be calculated with the same method.

Similarly to our previous paper, a conditional optimization has been performed, by taking into account that the *Purcell factor* and $QE$ of the coupled dipole – nanoshell resonator systems are inversely proportional [17]. The criterion set regarding the $\delta R_{excitation}^{rad}$ radiative rate enhancement at the excitation was that it has to be larger than unity, while a $cQE$ larger than a specific value, namely 50-40-30-20% was demanded at the emission. These $cQE$ values correspond to approximately 5-4-3-2-fold $\delta QE$ apparent quantum efficiency enhancement of SiV centers. The objective function was the product of radiative rate enhancements at the excitation and emission wavelengths, which is nominated as $P_x$ factor:

$$P_x = \delta R_{excitation}^{rad} \cdot \delta R_{emission}^{rad} \quad (3).$$

The in-house developed GLOBAL optimization algorithm was implemented into COMSOL [32, 33]. The configuration of four different coupled systems has been optimized:
1. Centralized spherical core-shell (CSCS): The dipole was exactly in the center of a spherical silver nanoshell and the fluorescence enhancement in the 2D parameter space of the $r_1$ core radius and $t$ shell thickness was inspected. One can expect only one single resonance peak in CSCS case in the inspected wavelength interval, because of the significant damping of the higher energy anti-bonding asymmetrical resonance.
2. Decentralized spherical core-shell (DSCS): A decentralized dipole is embedded into a spherical nanoresonator composed of a silver nanoshell. According to the symmetry breaking of illumination, multiple plasmonic resonances are expected, which can be precisely tuned to the desired wavelengths simultaneously.
3. Decentralized ellipsoidal core-shell (DECS): A decentralized dipole is embedded into an ellipsoidal core-shell nanoresonator composed of a silver nanoshell, which was elongated along one of its axes. Breaking the spherical symmetry of the concave nanoresonator inherently leads to multiple resonances, therefore it is expected that resonances with appropriate energy difference can be finely tuned to the desired wavelengths simultaneously.
4. Decentralized rod-like core-shell (DRCS): Deforming the ellipsoid into a rod-shaped nanoshell improves the antenna features of the concave nanoresontor, therefore it is expected that the optimized DRCS may lead to a larger radiative rate enhancement.

The far-field radiative rate enhancement spectra are presented for each type of core-shell resonators. To extract the wavelength dependent radiative rate enhancement, the frequency of the monochromatic dipole was swept in the [400 nm, 900 nm] interval. For further analyses the scattering cross-sections of nanoresonators were also determined via plane wave illumination in the same [400 nm, 900 nm] interval. The surface charge distribution on the nanoshell is also presented to uncover the type of underlying plasmonic resonances involved into the fluorescence enhancement. In addition to this, the spatial distribution of the **E**-field enhancement is presented as well. Although, in case of an experimental realization the parameters including the $r_1$ core radius and $t$ shell thickness, i.e. the $AR$, as well as the (x, y) dipole position and ($\varphi$, $\theta$) orientation have relatively large uncertainties, our results are presented with 0.1% accuracy, according to the high numerical stability of the finite element model computation. The optimized coupled SiV center - core-shell resonator systems are ranked based on the product of the $\delta R_{excitation}^{rad}$ excitation and $\delta R_{emission}^{rad}$ emission rate enhancements, namely on the $P_x$ factor. In the main text, we present the optimized coupled systems determined with 50% $cQE$ criterion, which have approximately 5-fold $\delta QE$ apparent quantum efficiency enhancement, whilst further optimization results are provided in the **Supplementary material**.

# 3. Results

## 3.1. Centralized dipole in optimized spherical concave core-shell nanoresonator

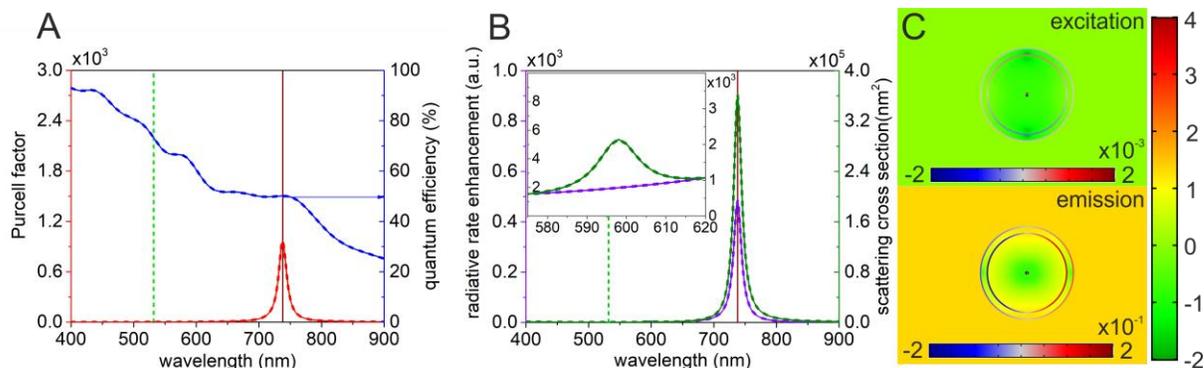

**Figure** 1. Optical response of an optimized spherical core-shell nanoresonator containing a centralized dipole. (A) Purcell factor i.e. total decay rate enhancement and quantum efficiency (B) radiative rate enhancement and scattering cross-section spectra of the optimized configuration corresponding to excitation (dashed lines) and emission (solid lines), inset: zoomed spectra around 600 nm. (C) Distribution of the surface charge density in arbitrary units and the normalized **E**-field enhancement with respect to vacuum on a logarithmic scale at the SiV excitation (top) and emission (bottom).

The simplest inspected configuration is, when a centralized dipole is embedded into a spherical core-shell nanoresonator closed by a silver nanoshell. Here only the $r_1$ core radius and $t$ shell thickness were varied and the optimization resulted in a core-shell with 0.854 aspect ratio ($AR=r_1/(r_1+t)$), which exhibits a resonance and corresponding *Purcell factor* peak exactly at the SiV emission (Fig. 1A and B). Based on that the spectral position of the plasmon resonance is dependent mainly on the aspect ratio of a core-shell structure close to the quasistatic limit, one could expect that by tuning the *AR* it is possible to tune the plasmonic resonance to the SiV excitation and hence to enhance the fluorescence via excitation rate enhancement (SFig. 1). However, caused by the criterion set during optimization regarding that the *cQE* must approach 50% at the emission, the optimization resulted in a coupled system exhibiting a large $\delta R^{rad}_{emission}$ radiative rate enhancement at SiV emission, which is in accordance with Eq (2).

The coupling of the centralized SiV color center to the spherical core-shell resonator results in large 482-fold emission rate enhancement, while the 1.1-fold radiative rate enhancement at the wavelength of excitation is only slightly larger than unity. The product of radiative rate enhancements is $P_x$=529 in case of the optimized CSCS (STable 1). An important advantage of the centralized spherical core-shell configuration is that the *QE* is high (73.2% and 49.7%) both at the excitation and emission wavelengths. Accordingly, the spherical core-shell nanoresonator enhances the intrinsic *QE* of the centralized SiV by a factor of 4.97. In case of this simple coupled system the results obtained by the in-house developed optimizing algorithm were verified by a parametric sweep above the $r_1$ core radius and $t$ shell thickness space (SFig. 1 and 2). It was also revealed that a gradually increasing *QE* can be obtained by increasing the core radii at the excitation wavelength, while there is a narrow optimal parameter region capable of maximizing the *cQE* at the emission. However, the parameter region appropriate to enhance the $\delta R^{rad}_{excitation/emission}$ radiative rate is governed by the *Purcell factor* at both wavelengths (SFig. 1 and 2).

Due to the rotational symmetry of the spherical core-shell nanoresonator the optical response is independent of the spatial orientation of the electric dipole, accordingly the optical responses take on the same values in excitation and emission configurations. Consequently, only dipolar plasmonic modes can be excited both at the excitation and emission wavelengths. Detailed inspection of the charge and near-field distribution shows that the dipolar charge distributions on the CSCS are perpendicular to each other, but the resonance is exclusively cavity and partially sphere plasmon like and the corresponding **E**-field enhancement is very weak and significantly stronger at the excitation and emission, respectively (Fig. 1C, top and bottom). At both wavelengths along the dipole oscillation direction a well-defined **E**-field depletion is observable around the dipole in the resonator and inside the shell with respect to that in a homogeneous environment. The near-field is the weakest inside the metal shell, where the strongest charge accumulation with the same sign is observable on the inner and outer interface, in accordance with the literature [5].

The near-field is enhanced along the dipole oscillation direction in the close vicinity of the metal shell due to the evanescent field of the LSPR, while the enhancement farther on outside perpendicularly to the dipole oscillation direction corresponds to the far-field radiation pattern of the coupled dipole-nanoresonator system. The strength of the **E**-field enhancement correlates with the $\delta R^{rad}_{excitation/emission}$ radiative rate enhancement values.

The selection rules of plasmon hybridization predict that two types of localized resonances could appear, however the charge distributions show that only the lower energy bonding resonance is at play in SiV fluorescence enhancement, while the higher energy anti-bonding resonance is out of the inspected wavelength interval (Fig. 1C). The absence of anti-bonding resonance is caused by the moderate charge separation, which is limited by the small size of the optimized spherical core-shell resonator. This explains that the *Purcell factor* and $\delta R^{rad}_{excitation/emission}$ radiative rate enhancement of the optimized CSCS configuration shows one single peak originating from bonding resonance causing that an efficient resonant coupling occurs only at the wavelength of emission (Fig.1A). Although, the scattering cross-section for plane wave illumination indicates a peak close to SiV excitation, the centralized dipole is not capable of resulting in a charge separation, which could promote radiation escaping at this wavelength.

## 3.2 Decentralized dipole in optimized spherical concave core-shell nanoresonator

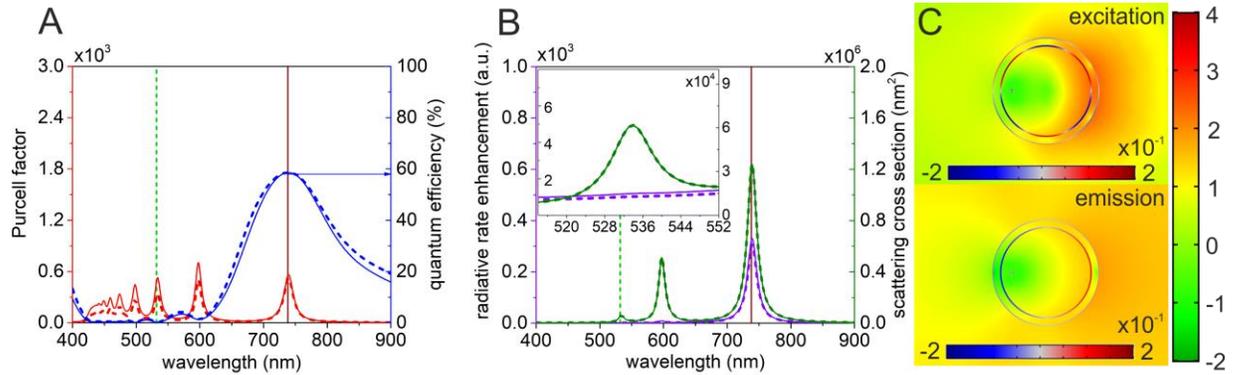

**Figure** 2. Optical response of an optimized spherical core-shell nanoresonator containing a decentralized dipole. (A) Purcell factor i.e. total decay rate enhancement and quantum efficiency (B) radiative rate enhancement and scattering cross section spectra of the optimized configuration corresponding to the excitation (dashed lines) and emission (solid lines), inset: zoomed spectra around the excitation wavelength. (C) Distribution of the surface charge density in arbitrary units and the normalized **E**-field enhancement with respect to vacuum on logarithmic scale at the SiV excitation (top) and emission (bottom).

In the second step, the dipole was decentralized inside a spherical concave core-shell nanoresonator. In this case one more variable has to be optimized, namely the δx displacement of the dipole from the center as well. The 0.852 *AR* of the optimized DSCS is very close to that of the optimized CSCS shown in Section 3.1. Decentralization by $\delta x$=29.6 nm results in that the optical response is no longer independent of the dipole spatial orientation. Accordingly, in excitation and emission configurations different far-field responses are collected, please note the differences between spectra determined via wavelength sweeps, indicated by dashed and solid curves in Fig. 2A, B.
Our results surprisingly show that dipole decentralization is not advantageous in case of a spherical core-shell resonator. Optimal configurations of DSCS always exhibit nearly centralized optimal dipole locations, however some local maxima were found in total fluorescence enhancement in case of a larger dislocations, such as the DSCS configuration presented here. In the optimized DSCS the 1.01 excitation rate enhancement is just above unity and is commensurate with that in CSCS, however it already fulfills the criterion regarding the excitation enhancement. The 319-fold emission rate enhancement is 2/3 times smaller than that in the centralized case, as a consequence the total fluorescence enhancement, namely the $P_x$ factor is 321-fold in DSCS. The *QE* is smaller at the excitation, than at the emission, i.e. their ratio is reversed with respect to that in CSCS. Despite the same 50% *cQE* criterion, the optimization results in a decentralized system with significantly better 57.5% *cQE,* in accordance with the larger $r_1$ dielectric core radius (STable 1.). The achieved 5.75-fold apparent SiV *QE* improvement also reveals that the forced dipole decentralization leads to local maxima in emission enhancement.
In DSCS the dipole is moved away the center of nanoshell, which breaks the spherical symmetry of the **E**-field with respect to the nanoshell center, moreover leads to higher order multipolar resonances at smaller wavelengths (Fig. 2A-C). Namely, at increasing energies a bonding quadrupolar, hexapolar and even higher order resonances appear.

Significantly different near-field phenomena are observable according to the different $\varphi$ oscillation directions of the dipoles corresponding to the transition moments of the SiV center, hence to their perpendicularity (Fig. 2C). In the presented optimal DSCS configuration a hexapolar mode enhances the excitation, however it is strongly nonradiative, as it is indicated on the wavelength dependent $QE$ quantum efficiency and $\delta R_{excitation}^{rad}$ radiative rate enhancement (Fig. 2A, B and C top). The optimal configuration exhibits a dipolar resonance at the SiV emission in case of DSCS similarly to CSCS (Fig. 2C bottom). Although, the decreased dipole distance from the metal could ensure a larger total decay rate, namely *Purcell factor* than in CSCS, decentralization leads to a significantly weaker dipole resonance of the coupled color-center and nanoresonator system at the SiV emission wavelength in DSCS, which manifest itself in a smaller $\delta R_{emission}^{rad}$ radiative rate enhancement.

Detailed inspection of the charge and near-field distribution shows that both the hexapolar charge distribution at the excitation and the dipolar charge distributions at the emission are more cavity plasmon like and the corresponding **E**-field enhancements are more commensurate than in CSCS (Fig. 2C). At both wavelengths, a well-defined **E**-field depletion is observable with respect to that in a homogeneous environment around the dipole in the resonator, along its oscillation direction. However, the near-field is the weakest inside the metal shell, where the distance is the smallest from the dipole. Further non-equidistantly distributed five minima (one opposing minimum) appear(s) at the SiV excitation (emission) wavelength, where the surface charge distribution has additional five local maxima (one local maximum). Considerable near-field enhancement is observable at the excitation as well, however its distribution is governed by the induced hexapole. The near-field enhancement of the coupled dipole-nanoresonator at both wavelengths is larger at the nanoresonator side opposing the dipole, caused by the broken symmetry of the illumination configuration. This can be explained by that the local strength of the LSPR on the core-shell nanoresonator almost equals, while the **E**-field of the reference dipole in a homogeneous environment is much weaker at the opposite side.

Although, the scattering cross-section for plane wave illumination indicates a larger peak at the SiV excitation, the nonradiative hexapolar charge separation even less promotes the radiation escaping at excitation, than the dipolar distribution in CSCS. In accordance with the spatially inhomogeneous near-field enhancement at the emission wavelength, and the smaller scattering cross-section for plane wave illumination, the induced dipole radiates less efficiently into the far-field with an oscillation center determined by the dipole dislocation in DSCS.

## 3.3 Decentralized dipole in optimized ellipsoidal concave core-shell nanoresonator

Symmetry breaking of the illumination configuration can be achieved not only by decentralizing the dipolar source inside a spherical nanoresonator, but also by deforming the shape of the concave core-shell nanoresonator itself. In present case deformation means the elongation of the spherical core-shell along the x-axis and creation of an ellipsoid. Nanoresonators with two different symmetry axes inherently have the potential to couple the light simultaneously into transversal and longitudinal plasmon resonances, which have smaller and larger resonance energies, respectively [18-20].

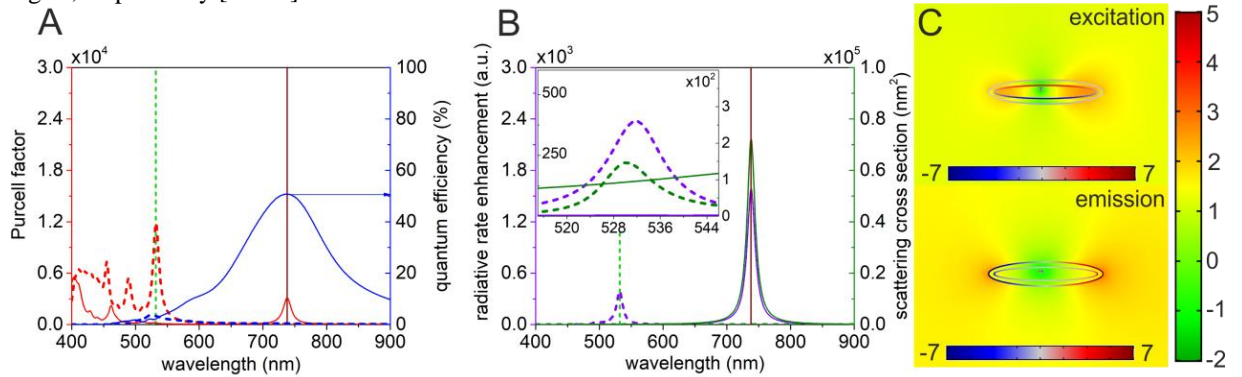

**Figure** 3. Optical response of an optimized ellipsoidal core-shell nanoresonator containing a decentralized dipole. (A) Purcell factor i.e. total decay rate enhancement and quantum efficiency (B) radiative rate enhancement and scattering cross-section spectra of the optimized configuration corresponding to the excitation (dashed lines) and emission (solid lines), inset: zoomed spectra around the excitation wavelength. (C) Distribution of the surface charge density in arbitrary units and the normalized **E**-field enhancement with respect to vacuum on logarithmic scale at the excitation and emission.

Elongation of a core-shell nanoresonator has multiple effects on the optical response (Fig. 3). Most important is the splitting of the LSPR into longitudinal and transversal modes, which already occurs in case of a centralized dipole. Further effect of the core-shell nanoresonator elongation is the reduction of the dipole distance from the metal shell along the short axes. The shorter the distance is, the higher is the total decay rate enhancement qualified by the *Purcell factor*. However, this can cause a decrease in quantum efficiency according to the increased heat dissipation into the metal that occurs in case of decreased distances.

Similarly to the case of decentralized spherical core-shell nanoresonators, DECS has multiple higher order resonances at smaller wavelengths (Fig. 3A), which correspond to the transversal and longitudinal resonances of the concave metal ellipsoid. Most of them are dark modes with very low quantum efficiency and hence have insignificant radiative rate enhancement at their resonance wavelengths. In case of ($\delta x$=-4.29 nm, $\delta y$=2.37 nm) decentralization, 4.73 axis ratio and 4.76 nm shell thickness a resonance and a corresponding *Purcell factor* peak appears at the SiV excitation and emission in the corresponding configurations (Fig. 3B). As a result, in the optimized DECS well defined peak appears as well as on the wavelength dependent $\delta R_{excitation/emission}^{rad}$ radiative rate enhancement in the corresponding configurations. Although, the *QE* is only 3.3% at the excitation wavelength, the resonance strength is almost two orders of magnitude higher than that of the hexapolar mode in DSCS, which results in an excitation rate enhancement of 392. Despite the small 3.32 nm dipole distance from metal the resonance at the emission exhibits 50.6% *cQE* with a stronger total decay rate enhancement. The $1.58 \cdot 10^3$ $\delta R_{emission}^{rad}$ emission rate enhancement is one order of magnitude larger than in CSCS and DSCS, revealing that the ellipsoidal core-shell nanoresonator exhibits a plasmonic resonance, which is capable of resulting in a more significant radiation escaping, than the spherical ones (Fig. 3A, B). The total fluorescence enhancement, namely the $P_x$ factor is $6.2 \cdot 10^5$-fold in the optimized DECS configuration.

The dipole moments lie along the short- and long semi axes at the excitation and emission, respectively, which ensure the full exploitation of the transition dipole moment orthogonality. The simultaneous excitation and emission enhancement is due to the higher (lower) energy resonance at the excitation (emission), which originates from a transversal (longitudinal) dipolar mode along the short (long) axis of the ellipsoid according to the surface charge distribution (Fig. 3C top (bottom)). The charge distribution indicates that the cavity plasmon is dominant at the first order transversal resonance. This resonance is less radiative because of the higher material loss in the metal at smaller wavelength, the small dipole distance and the significant contribution of the primitive cavity plasmon (Fig. 3C top). In contrast, the lower energy resonance corresponds to the longitudinal dipolar resonance, and the charge distribution indicates that the primitive spheroid plasmon is dominant in this case (Fig. 3C bottom).

The near-field enhancement is higher inside and outside the nanoellipsoid at the excitation and emission, respectively (Fig. 3C). At both wavelengths, a well-defined **E**-field depletion with respect to that in a homogeneous environment is observable in the resonator around the dipole, which is oriented along the short axis at both wavelengths however it is more confined at the excitation. Surprisingly, the near-field is the weakest along the short axes as well as inside the metal shell independently of the dipole orientation, even if the surface charge distribution has local maxima and minima at the excitation and emission wavelength, respectively. At the excitation wavelength, the near-field of the coupled dipole-nanoresonator exhibits the strongest enhancement inside the nanoresonator between the oppositely charged walls in accordance with the literature [5]. The enhancement is not homogeneous, it is larger along the long axis of the ellipsoidal nanoresonator, while it is relatively smaller at the resonator walls adjacent to the dipole. In contrast at the emission wavelength the near-field enhancement originating from longitudinal resonance is concentrated outside the nanoresonator, moreover at the tips of the nanorods the local enhancement reveals to lightening rod effect.

It is important to note that even though the scattering efficiency of the DECS is significantly smaller than that of CSCS and DSCS at the SiV excitation, a two orders of magnitude larger $\delta R_{excitation}^{rad}$ radiative rate enhancement occurs, indicating that the ellipsoid could promote light escaping also at this wavelength. Despite the commensurate near-field enhancements, dominantly the induced longitudinal dipole radiates efficiently into the far-field, again with an oscillation center governed by the dipole dislocation.

## 3.4 Decentralized dipole in optimized rod-shaped concave core-shell nanoresonators

We have optimized also a rod-shaped concave core-shell nanoresonators, where higher $\delta R^{rad}_{excitation/emission}$ radiative rate enhancements are expected according to their more robust antenna properties. In many aspects, the optimized rod-shaped core-shell nanoresonator is similar to the optimized DECS presented in Section 3.3. In case of ($\delta x$=0 nm, $\delta y$=1·10$^{-2}$ nm) decentralization, 4.4 axis ratio and 4.55 nm shell thickness resonance peak appears at the SiV excitation and emission in the corresponding configuration, moreover multiple resonances develop below the SiV excitation wavelength, despite the significantly less decentralized dipole source with respect to DECS (Fig. 4A).

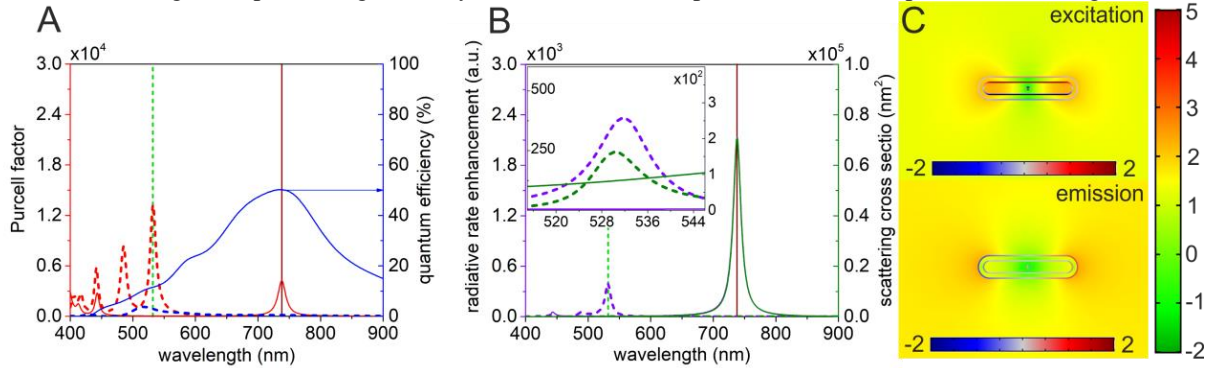

**Figure** 4. Optical response of an optimized rod-shaped core-shell nanoresonator containing a decentralized dipole. (A) Purcell factor i.e. total decay rate enhancement and quantum efficiency (B) radiative rate enhancement and scattering cross section spectra of the optimized configurations corresponding to the excitation (dashed lines) and emission (solid lines), inset: zoomed spectra around the excitation wavelength. (C) Distribution of the surface charge density and the normalized **E**-field enhancement with respect to vacuum on logarithmic scale at the excitation and emission.

Although, the resonances in DRCS are stronger than in DECS, the slightly smaller 2.94% *QE* makes it possible to reach a 388-fold excitation rate enhancement, which is slightly smaller than in DECS. There is a more significant difference between DECS and DRCS at the SiV emission. Namely, the longitudinal dipole resonance tuned to the SiV esmission nm is accompanied by 4/3-times stronger 2150 radiative rate enhancement in DRCS, while the 50.3% *cQE* is slightly smaller. The excitation and emission enhancements together lead to a 4/3-times stronger $P_x$ factor, since 8.34·10$^5$ total fluorescence enhancement is achieved.

Again, only the dipolar transversal and longitudinal resonances contribute significantly to the $\delta R^{rad}_{excitation/emission}$ radiative rate enhancements at the excitation and emission wavelengths due to their relatively large *QE*. The rod's geometry is capable of enhancing the excitation via the transversal dipolar cavity mode, which is a bonding mode according to surface charge distribution (Fig. 4C top). In contrast, the longitudinal dipolar spheroid mode, which is again a bonding mode, makes possible the most significant emission enhancement among the inspected concave plasmonic resonators.

The near-field enhancement is higher inside and outside the nanorod at the excitation and emission, respectively (Fig. 4C). At both wavelengths a well-defined **E**-field depletion is observable with respect to that in a homogeneous environment around the dipole in the resonator, however it is oriented again along the short axis and it is more confined at the excitation, similarly to the optimized DECS. Inside the metal shell the near-field is the weakest along the short axes at the excitation, while at the emission wavelength there are similarly weak near-field areas along the short (long) axes, where the surface charge distribution has local minima (maxima). At the SiV excitation wavelength, the near-field of the coupled system exhibits the strongest enhancement inside the nanoresonator between the oppositely charged walls, similarly to the nanoellipsoid [5]. The enhancement is not homogeneous, it is again larger along the long axis of the nanorod, similarly to DECS. In contrast, at the SiV emission wavelength near-field enhancement originating from the longitudinal resonance is observable exclusively outside the nanoresonator.

Despite the commensurate near-field enhancements, the induced longitudinal dipole radiates one order of magnitude more efficiently into the far-field with respect to the transversal dipole. The near-field enhancement is more symmetric at both wavelengths with respect to the DECS due to the almost centralized dipole. The stronger resonance at the emission is accompanied by a slightly smaller *cQE*, which indicates that a more antenna-like rod-shaped core-shell nanoresonator is capable of promoting the radiation escape despite the scattering efficiency commensurate with that of the optimized DECS consisting of a gradually narrowing diamond core.

# Conclusions

In conclusion we have shown that significant fluorescence enhancement is achievable by optimizing all of the inspected concave plasmonic core-shell nanoresonators. The optimized CSCS configuration is capable of resulting in a 428-fold emission enhancement due to the dipolar bonding LSPR tuned to emission wavelength, whilst at the excitation no significant enhancement occurs because of the absence of higher order multipolar modes and the out of range antibonding resonance. The $P_x$ *factor* of CSCS is 529 with a 49.7% *cQE* at the emission, which corresponds to an approximately 5-fold *δQE* apparent quantum efficiency enhancement.

We have found that decentralization of the emitter is not advantageous, since optimizations always result in a global maximum corresponding to a centralized dipole. In case of a local maximum corresponding to the inspected DSCS, due to the broken symmetry of illumination, higher order multipolar modes appear on the enhancement spectra. Although, in this case the hexapolar bonding resonance is tuned to the excitation, no significant excitation enhancement occurs caused by the nonradiative nature of this higher order multipolar mode. Similarly to the optimized CSCS, the optimized DSCS also has its dipolar resonance at the emission wavelength, which results in a 319-fold emission rate enhancement and a 321 $P_x$ *factor*. These parameters are smaller than in CSCS, however the advantage of DSCS is that the fluorescence enhancement is accompanied by a larger 57.5% *cQE*.

In case of the optimized DECS higher order multipolar modes also appear at higher energies, however the separation of transversal and longitudinal LSPR due to different short and long axis lengths makes it possible to tune bright dipolar modes to the excitation and emission wavelengths simultaneously. In the optimized DECS possessing the transversal dipolar mode in the excitation and the longitudinal dipolar mode in the emission configuration the $P_x$ *factor* is $6.20 \cdot 10^5$, which is more than three-orders of magnitude larger than in CSCS. This large $P_x$ *factor* is accompanied by the intermediate 50.6% *cQE*, which corresponds to an approximately 5-fold *δQE* enhancement.

The optimized DRCS exploits transversal and longitudinal dipolar resonances to enhance SiV fluorescence similar to those observed in the optimized DECS configuration according to their similar shape, manifesting itself in similar long-to-short axes ratio. However, DRCS shows a slightly smaller excitation rate enhancement than DECS, while its radiative rate enhancement at the emission is almost 1.5-times larger according to the more robust antenna properties, which is accompanied by a similar intermediate 50.3% *cQE* apparent quantum efficiency. The $P_x$ *factor* of the optimized rod-shaped concave core-shell nanoresonator is $8.34 \cdot 10^5$, which indicates is the highest total fluorescence enhancement among the presented nanoresonators accompanied by 5-fold *δQE*. It is possible to increase the enhancement even higher, however it results in slightly smaller *cQE* hence allows smaller apparent quantum efficiency enhancement.

# Improved emission of SiV diamond color centers embedded into concave plasmonic core-shell nanoresonators

András Szenes[1], Balázs Bánhelyi[2], Lóránt Zs. Szabó[1],
Gábor Szabó[1], Tibor Csendes[2], Mária Csete[1,*]

[1]Department of Optics and Quantum Electronics, University of Szeged, Dóm tér 9, Szeged, H-6720, Hungary.
[2]Institute of Informatics, University of Szeged, Árpád tér 2, Szeged, H-6720, Hungary

*mcsete@physx.u-szeged.hu, +36-62-544654

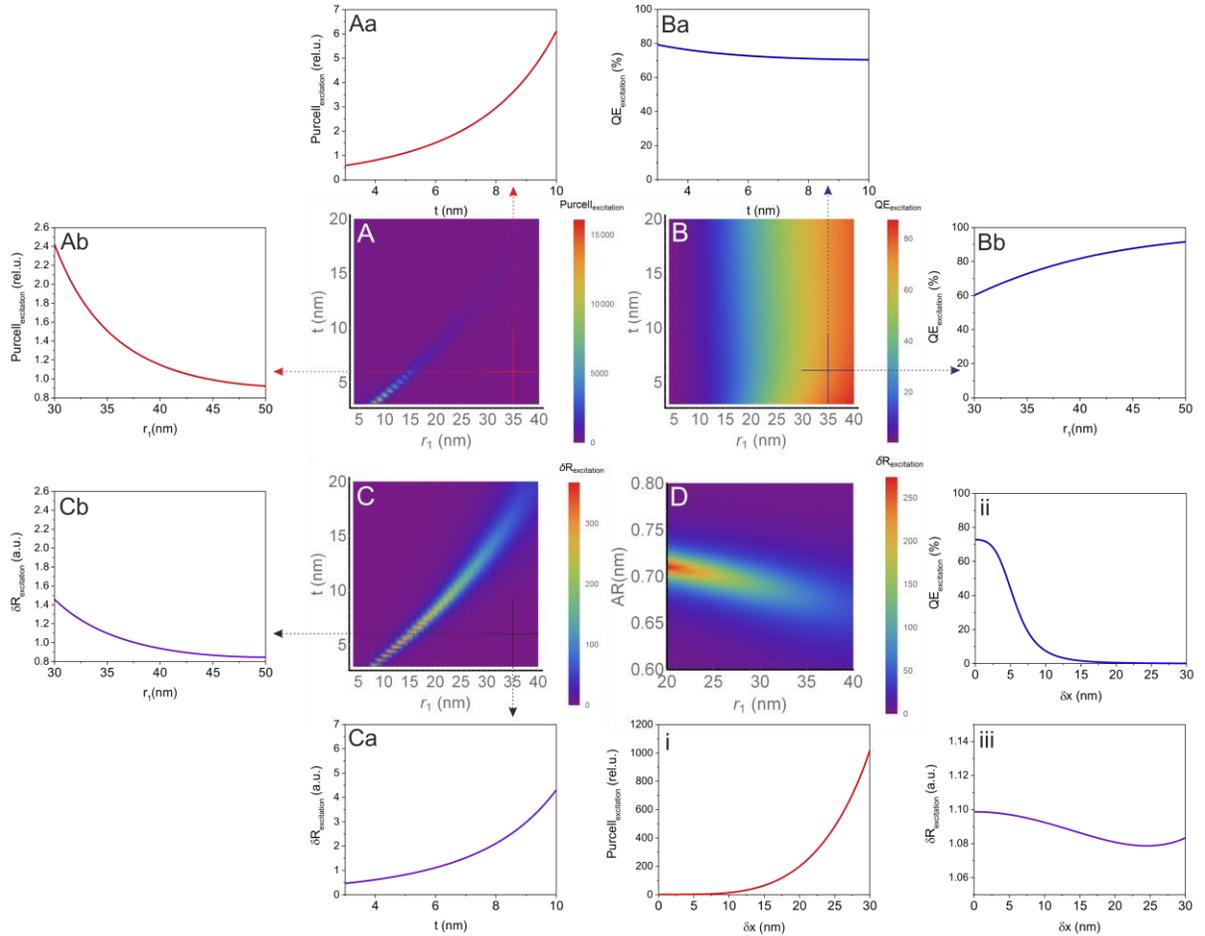

**SFigure 1**. Optical response of CSCS configuration above the core radius ($r_1$) – shell thickness ($t$) parameter space. (A) *Purcell factor* (B) *QE* (C) $\delta R^{rad}_{excitation}$ radiative rate enhancement vs. core radius and shell thickness (D) $\delta R^{rad}_{excitation}$ radiative rate enhancement vs. core radius and *AR* at SiV excitation wavelength. Insets: projections of the 2D parameter space (Aa) *Purcell factor* vs. shell thickness at optimal (35 nm) core radius and (Ab) *Purcell factor* vs. core radius at optimal (5.96 nm) shell thickness, (Ba) *QE* vs. shell thickness at optimal (35 nm) core radius and (Bb) *QE* vs. core radius at optimal (5.96 nm) shell thickness, (Ca) $\delta R^{rad}_{excitation}$ radiative rate enhancement vs. core radius at optimal (5.96 nm) shell thickness and (Cb) $\delta R^{rad}_{excitation}$ radiative rate enhancement vs. shell thickness at optimal (35 nm) core radius at 532 nm. (i-iii) *Purcell factor*, *QE*, $\delta R^{rad}_{excitation}$ radiative rate enhancement vs. dipole displacement at SiV excitation wavelength.

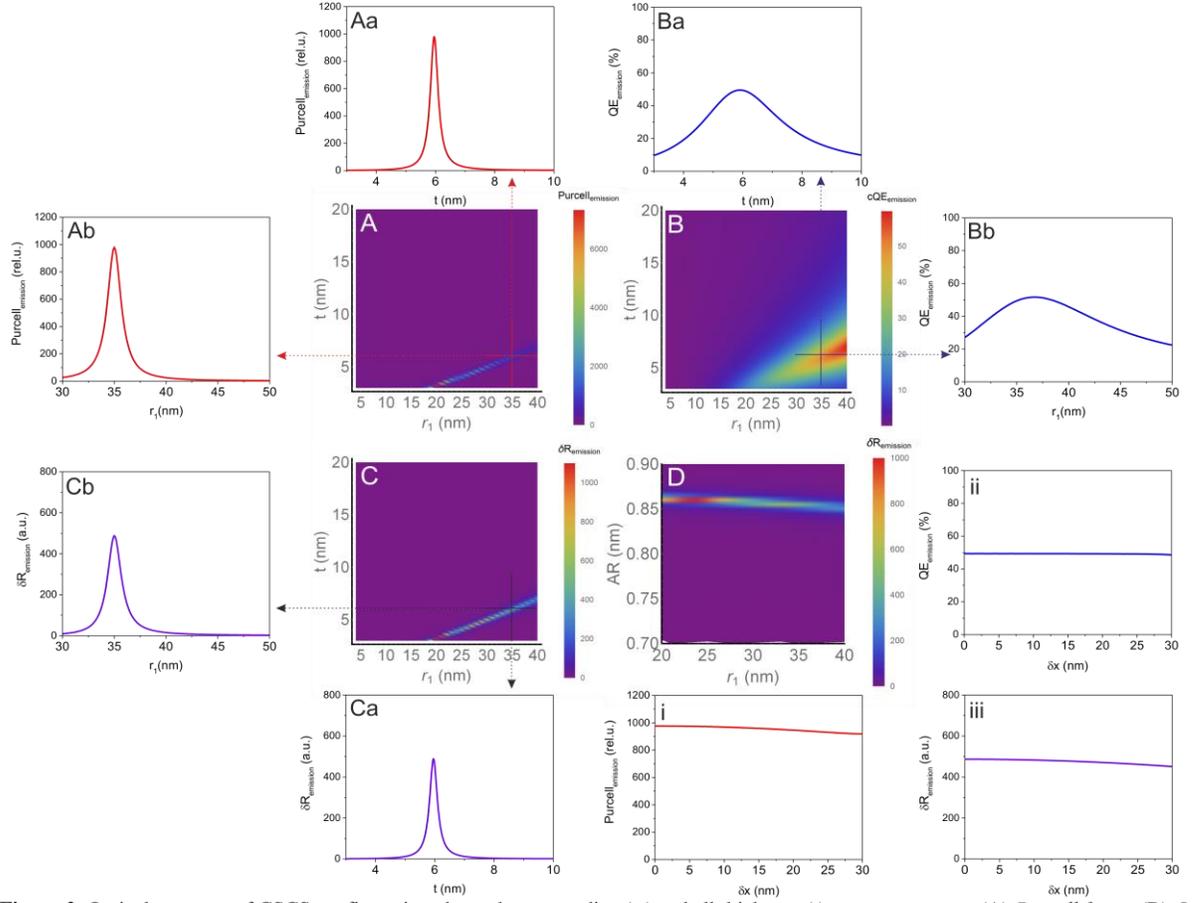

**SFigure 2**. Optical response of CSCS configuration above the core radius ($r_1$) – shell thickness ($t$) parameter space. (A) *Purcell factor* (B) *QE* (C) $\delta R^{rad}_{emission}$ radiative rate enhancement vs. core radius and shell thickness, (D) $\delta R^{rad}_{emission}$ radiative rate enhancement vs. core radius and *AR* at SiV emission wavelength. Insets: projections of the 2D parameter space (Aa) *Purcell factor* vs. shell thickness at optimal (35 nm) core radius and (Ab) *Purcell factor* vs. core radius at optimal (5.96 nm) shell thickness, (Ba) *QE* vs. shell thickness at optimal (35 nm) core radius and (Bb) *QE* vs. core radius at optimal (5.96 nm) shell thickness, (Ca) $\delta R^{rad}_{emission}$ radiative rate enhancement vs. core radius at optimal (5.96 nm) shell thickness and (Cb) $\delta R^{rad}_{emission}$ radiative rate enhancement vs. shell thickness at optimal (35 nm) core radius at 738 nm. (i-iii) *Purcell factor*, *QE*, $\delta R^{rad}_{emission}$ radiative rate enhancement vs. dipole deposition at SiV emission wavelength.

Only one single resonance peak appears in the inspected geometrical parameter interval both at the SiV excitation and at the SiV emission wavelengths (SFig. 1A and SFig. 2A). The resonance energy is dependent mainly (exclusively) on the aspect ratio at the excitation (emission) (SFig. 1D and 2D), only slight deflection is observable on the curve corresponding to the $\delta R^{rad}_{excitation/emission}$ radiative rate enhancement above the *AR* aspect ratio and $r_1$ core radius parameter space. The *cQE* corrected quantum efficiency can be high only at radiative resonances (SFig. 2B).

# Comparative study on coupled SiV color center - concave nanoresonator systems optimized with different criteria

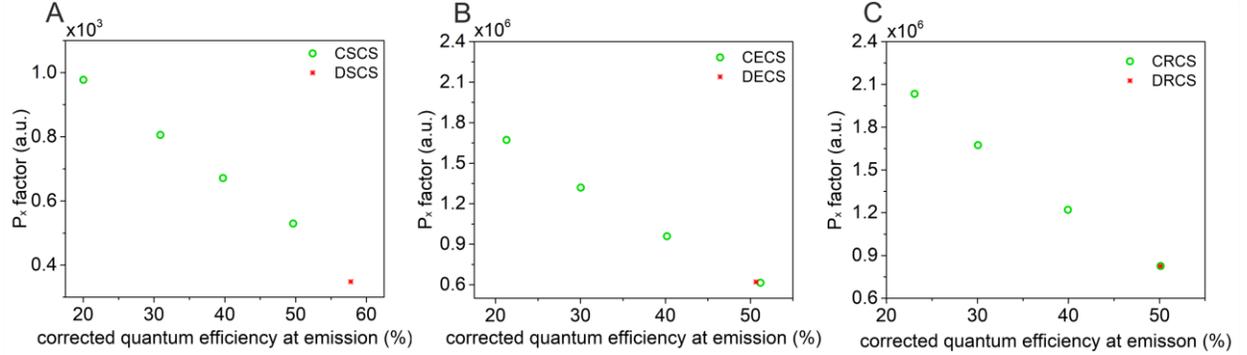

**SFigure 3** Total fluorescence enhancement ($P_x$ *factor*) of (A) spherical, (B) ellipsoidal and (C) rod-shaped concave core-shell nanoresonators containing a centralized dipole and optimized with different *cQE* criteria. The $P_x$ *factor* of the optimized configurations consisting of a decentralized dipole corresponding to *cQE*=50% are also shown by red stars.

The optimization results show that in all optimized configurations at the emission wavelength the $\delta R_{emission}^{rad}$ radiative rate enhancement of SiV color center decreases by increasing the criterion regarding the corrected quantum efficiency.

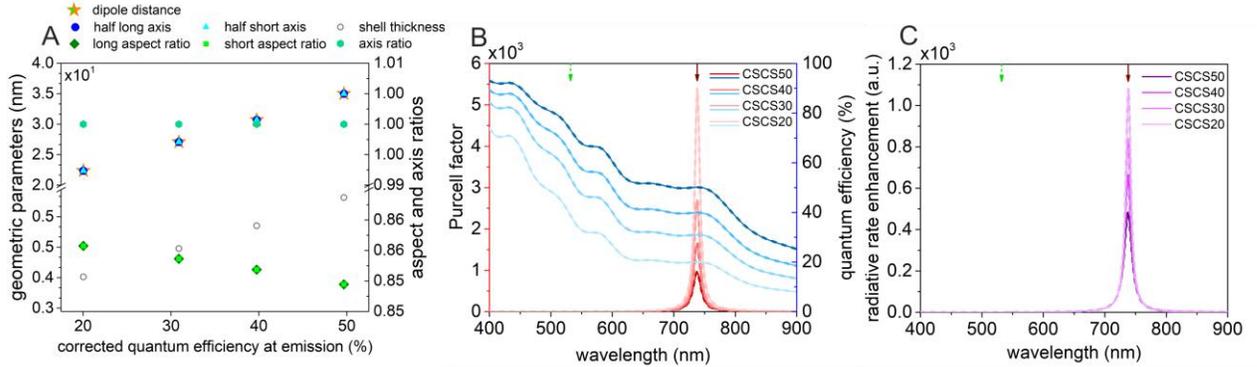

**SFigure 4** Parameters and optical response of CSCS optimized with 50, 40, 30 and 20% *cQE* criteria.

In case of CSCS the achievable *cQE* is governed by the distance between the emitter and the absorbing metal, as well as by the metal shell thickness. Namely, the higher the desired *cQE* is, the larger is the diamond core radius, and surprisingly the *t* shell thickness increases as well. As a result, the total spherical core-shell nanoresonator size increases nearly linearly in the quasistatic limit by increasing the *cQE* criterion (SFig. 4A).

The position of the sole visible resonance peak is mainly determined by the *AR* aspect ratio of the concave core-shell nanoresonator (STable 1.). In all optimized coupled systems the symmetric bonding dipolar mode is tuned to 738 nm. Accordingly, the *AR* of nanoshells are nearly the same, only a slight *AR* decrease is observable as the *cQE* criterion increases (SFig. 4A). However, the larger core-shell nanoresonator size leads to a weaker resonance as the *Purcell factor* spectra show (SFig. 4B). As a consequence, the $\delta R_{emission}^{rad}$ radiative rate enhancement at the emission decreases by increasing the *cQE* criterion (SFig. 4C).

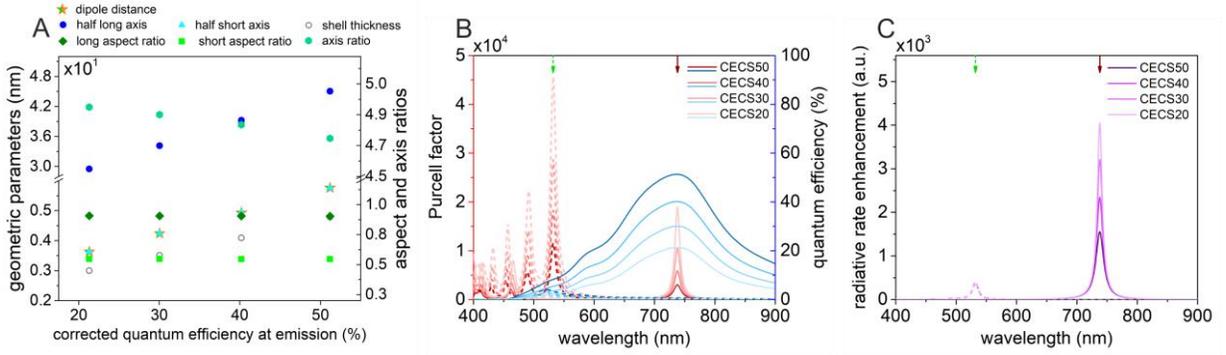

**SFigure 5** Parameters and optical response of CECS optimized with 50, 40 30 and 20% *cQE* criteria.

The interdependence of the geometrical parameters and the resonance energy is more complex in case of concave ellipsoidal core-shell nanoresonators. Generally, it can be concluded again that larger *cQE* requires larger emitter-metal distance (Fig. 5A). To increase the dipole distance one should increase the short axis of the ellipsoid as well. To maintain the transversal and longitudinal resonances at SiV excitation and emission wavelengths, the ratio of long and short axes should be conserved. Accordingly, the higher the desired *cQE* is, the larger is the diamond core short and long axis, and surprisingly the *t* shell thickness increases as well. As a result, the total ellipsoidal core-shell nanoresonator size increases similarly to CSCS.

In the *ARs* corresponding to the short and long axis of nanoshells only a slight decrease is observable, as the *cQE* criterion increases (STable 1). Moreover, not only the aspect ratio but the axes lengths themselves influence the energies of the resonance peaks. The larger *cQE* at emission requires larger metal-emitter distance, accordingly larger short and long axis. A slightly smaller long-to-short axis ratio is preferred, as the *cQE* increases.

Similarly to CSCS nanoresonators, the higher is the *cQE*, the lower is the *Purcell factor* and the $\delta R^{rad}_{excitation/emission}$ radiative decay rate enhancement both at the excitation and emission (SFig. 5 B, C). In CECS by decreasing the *cQE* criterion from 50% through 20%, the relative radiative rate enhancement increase is insignificant at the excitation, while at the emission the increase is more than 2-fold.

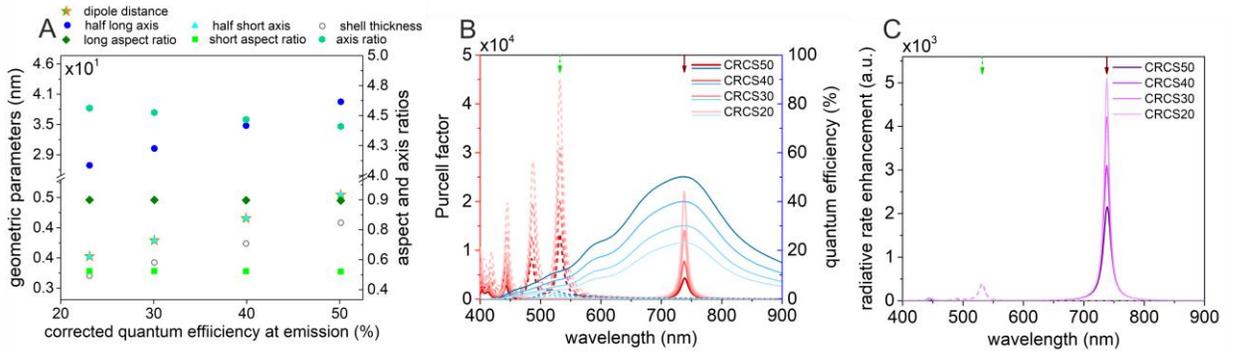

**SFigure 6** Parameters and optical response of CRCS optimized with 50, 40, 30 and 20% *cQE* criteria.

In case of CRDS the tendencies are the same as in CECS, namely to achieve higher *cQE* larger dipole distance is required, which implies appropriatley larger short and long axis, and maintainance of the optimal decreasing axis lenght ratio (SFig. 6A). Surprisingly the thickness of the metal shell increases as well. In the *ARs* corresponding to the short and long axis of nanoshells only a slight decrease is observable. The rod-shaped concave core-shell nanoresonators show a *Purcell factor* and $\delta R^{rad}_{excitation/emission}$ radiative rate enhancement characteristics and *QE* spectra similar to those of CECS (SFig. 6B, C). Although, their excitation rate enhancements are almost identical, a major difference between CRCS and CECS is that the CRCSs have more significantly larger *Purcell factor* at the emission. Despite the smaller dipole distance from the metal, at the same *cQE* larger $\delta R^{rad}_{emission}$ radiative rate enhancement is achieved in case of CRCSs, due to the efficient far-field coupling via rod-shaped nano-antennas. The highest $2.03 \cdot 10^6$ fluorescence enhancement among the inspected concave plasmonic nanoresonators was achieved via rod-shaped cores embedded into silver nanoshell, when the criterion regarding the minimum *cQE* was set to 20%.

|  | excitation | | | | emission | | | | | geometry | | | | | | | | |
|---|---|---|---|---|---|---|---|---|---|---|---|---|---|---|---|---|---|---|
|  | criterion* | Purcell factor | QE (%) | δR (a.u.)* | criterion* | Purcell factor | cQE (%)* | δR$^{rad}$ (a.u.) | Px factor (a.u.) | d (nm) | $r_1$ (nm) | $r_2$ (nm) | t (nm) | $AR_1$ | $AR_2$ | AXR | δx (nm) | δy (nm) |
| DSCS50 | 1 | 3.13E+02 | 3.22E-01 | 1.01E+00 | 50 | 5.55E+02 | 5.75E+01 | 3.19E+02 | 3.21E+02 | 9.59E+00 | 3.92E+01 | 3.92E+01 | 6.80E+00 | 8.52E-01 | 8.52E-01 | 1.00E+00 | 2.96E+01 | 0.00E+00 |
| CSCS50 | 1 | 1.50E+00 | 7.32E+01 | 1.10E+00 | 50 | 9.71E+02 | 4.97E+01 | 4.82E+02 | 5.29E+02 | 3.50E+01 | 3.50E+01 | 3.50E+01 | 5.96E+00 | 8.54E-01 | 8.54E-01 | 1.00E+00 | 0.00E+00 | 0.00E+00 |
| CSCS40 | 1 | 1.60E+00 | 6.37E+01 | 1.02E+00 | 40 | 1.66E+03 | 3.98E+01 | 6.60E+02 | 6.71E+02 | 3.07E+01 | 3.07E+01 | 3.07E+01 | 5.13E+00 | 8.57E-01 | 8.57E-01 | 1.00E+00 | 0.00E+00 | 0.00E+00 |
| CSCS30 | 0 | 1.79E+00 | 5.38E+01 | 9.64E-01 | 30 | 2.71E+03 | 3.09E+01 | 8.36E+02 | 8.06E+02 | 2.71E+01 | 2.71E+01 | 2.71E+01 | 4.46E+00 | 8.59E-01 | 8.59E-01 | 1.00E+00 | 0.00E+00 | 0.00E+00 |
| CSCS20 | 0 | 2.34E+00 | 3.87E+01 | 9.03E-01 | 20 | 5.40E+03 | 2.00E+01 | 1.08E+03 | 9.78E+02 | 2.24E+01 | 2.24E+01 | 2.24E+01 | 3.62E+00 | 8.61E-01 | 8.61E-01 | 1.00E+00 | 0.00E+00 | 0.00E+00 |

|  | excitation | | | | emission | | | | | geometry | | | | | | | | |
|---|---|---|---|---|---|---|---|---|---|---|---|---|---|---|---|---|---|---|
|  | criterion* | Purcell factor | QE (%) | δR (a.u.)* | criterion* | Purcell factor | cQE (%)* | δR$^{rad}$ (a.u.) | Px factor (a.u.) | d (nm) | $r_1$ (nm) | $r_2$ (nm) | t (nm) | $AR_1$ | $AR_2$ | AXR | δx (nm) | δy (nm) |
| DECS50 | 1 | 1.19E+04 | 3.30E+00 | 3.92E+02 | 50 | 3.13E+03 | 5.06E+01 | 1.58E+03 | 6.20E+05 | 3.32E+00 | 4.48E+01 | 5.72E+00 | 4.76E+00 | 9.04E-01 | 5.46E-01 | 4.73E+00 | 4.29E+00 | -2.37E+00 |
| CECS50 | 1 | 1.16E+04 | 3.42E+00 | 3.96E+02 | 50 | 3.03E+03 | 5.12E+01 | 1.55E+03 | 6.15E+05 | 5.76E+00 | 4.51E+01 | 5.76E+00 | 4.79E+00 | 9.04E-01 | 5.46E-01 | 4.72E+00 | 0.00E+00 | 0.00E+00 |
| CECS40 | 1 | 1.84E+04 | 2.21E+00 | 4.06E+02 | 40 | 5.86E+03 | 4.01E+01 | 2.35E+03 | 9.54E+05 | 4.93E+00 | 3.92E+01 | 4.93E+00 | 4.09E+00 | 9.06E-01 | 5.46E-01 | 4.80E+00 | 0.00E+00 | 0.00E+00 |
| CECS30 | 1 | 2.87E+04 | 1.43E+00 | 4.11E+02 | 30 | 1.07E+04 | 3.00E+01 | 3.21E+03 | 1.32E+06 | 4.24E+00 | 3.42E+01 | 4.24E+00 | 3.51E+00 | 9.07E-01 | 5.47E-01 | 4.86E+00 | 0.00E+00 | 0.00E+00 |
| CECS20 | 1 | 4.57E+04 | 9.04E-01 | 4.13E+02 | 20 | 1.90E+04 | 2.13E+01 | 4.05E+03 | 1.67E+06 | 3.63E+00 | 2.95E+01 | 3.63E+00 | 3.00E+00 | 9.08E-01 | 5.47E-01 | 4.90E+00 | 0.00E+00 | 0.00E+00 |

|  | excitation | | | | emission | | | | | geometry | | | | | | | | |
|---|---|---|---|---|---|---|---|---|---|---|---|---|---|---|---|---|---|---|
|  | criterion* | Purcell factor | QE (%) | δR (a.u.)* | criterion* | Purcell factor | cQE (%)* | δR$^{rad}$ (a.u.) | Px factor (a.u.) | d (nm) | $r_1$ (nm) | $r_2$ (nm) | t (nm) | $AR_1$ | $AR_2$ | AXR | δx (nm) | δy (nm) |
| DRCS50 | 1 | 1.32E+04 | 2.94E+00 | 3.88E+02 | 50 | 4.27E+03 | 5.03E+01 | 2.15E+03 | 8.34E+05 | 5.37E+00 | 3.92E+01 | 5.38E+00 | 4.55E+00 | 8.96E-01 | 5.42E-01 | 4.40E+00 | 0.00E+00 | 1.00E-02 |
| CRCS50 | 1 | 1.34E+04 | 2.90E+00 | 3.88E+02 | 50 | 4.27E+03 | 5.00E+01 | 2.14E+03 | 8.29E+05 | 5.36E+00 | 3.92E+01 | 5.36E+00 | 4.55E+00 | 8.96E-01 | 5.41E-01 | 4.41E+00 | 0.00E+00 | 0.00E+00 |
| CRCS40 | 1 | 2.01E+04 | 1.96E+00 | 3.93E+02 | 40 | 7.77E+03 | 3.99E+01 | 3.10E+03 | 1.22E+06 | 4.67E+00 | 3.46E+01 | 4.67E+00 | 3.94E+00 | 8.98E-01 | 5.43E-01 | 4.47E+00 | 0.00E+00 | 0.00E+00 |
| CRCS30 | 1 | 3.11E+04 | 1.28E+00 | 3.97E+02 | 30 | 1.40E+04 | 3.01E+01 | 4.22E+03 | 1.67E+06 | 4.03E+00 | 3.02E+01 | 4.03E+00 | 3.39E+00 | 8.99E-01 | 5.43E-01 | 4.53E+00 | 0.00E+00 | 0.00E+00 |
| CRCS20 | 1 | 4.48E+04 | 8.91E-01 | 3.99E+02 | 20 | 2.21E+04 | 2.31E+01 | 5.10E+03 | 2.03E+06 | 3.57E+00 | 2.70E+01 | 3.57E+00 | 3.00E+00 | 9.00E-01 | 5.43E-01 | 4.56E+00 | 0.00E+00 | 0.00E+00 |

**Stable 1**. Optical response and geometric parameters of optimized SCS, ECS and RCS configurations.

QE: quantum efficiency; $\delta R^{rad}_{excitation/emission}$: radiative rate enhancement; cQE: corrected quantum efficiency; $P_x$ *factor*: total fluorescence rate enhancement ($\delta R^{rad}_{excitation} \cdot \delta R^{rad}_{emission}$); d: minimum dipole distance from metal nanoshell; $r_1$: radius of diamond core along the long axis; $r_2$: radius of diamond core along the short axis; t: silver shell thickness; $AR_1$; generalized aspect ratio along the long axis ($r_1/(r_1+t)$); $AR_2$: generalized aspect ratio along the long axis ($r_2/(r_2+t)$); AXR: axis ratio (($r_1+t)/(r_2+t)$); δx: dipole decentralization along *x* axis; δy: dipole decentralization along *y* axis.